\documentclass[3p,times,twocolumn]{elsarticle}

\usepackage{ecrc}

\volume{00}

\firstpage{1}

\journalname{Nuclear Physics B Proceedings Supplement}

\runauth{}

\jid{nppp}

\jnltitlelogo{Nuclear Physics B Proceedings Supplement}

\usepackage{amssymb}
\usepackage[figuresright]{rotating}

\usepackage{amsmath,amssymb}
\usepackage{wasysym}
\usepackage{slashed}
\usepackage{graphicx}
\usepackage{pdfcomment}
\newcommand{\be}{\begin{equation}}
\newcommand{\ee}{\end{equation}}
\newcommand{\bea}{\begin{eqnarray}}
\newcommand{\eea}{\end{eqnarray}}
\newcommand{\f}{\frac}

\newcommand{\bra}{\langle}
\newcommand{\ket}{\rangle}

\begin{document}

\begin{frontmatter}



\dochead{}

\title{Radiative energy loss and radiative $p_\perp$-broadening of high-energy partons in QCD matter}
\author{Bin Wu}
\ead{Bin.WU@cea.fr}
\address{Institut de Physique Th\'{e}orique, CEA Saclay, 91191, Gif-sur-Yvette Cedex, France}

\begin{abstract}
I give a self-contained review on radiative $p_\perp$-broadening and radiative energy loss of high-energy partons in QCD matter. The typical $p_\perp^2$ of high-energy partons receives a double logarithmic correction due to the recoiling effect of medium-induced gluon radiation. Such a double logarithmic term, averaged over the path length of the partons, can be taken as the radiative correction to the jet quenching parameter $\hat{q}$ and hence contributes to radiative energy loss. This has also been confirmed by detailed calculations of energy loss by radiating two gluons.
\end{abstract}


\end{frontmatter}

% \linenumbers

\section{Introduction}
\label{sec:intro}
High energy partons serve as hard probes to bulk QCD matter created in ultra-relativistic heavy-ion collisions at RHIC and the LHC. The motion of such hard probes, determined by the properties of bulk matter, can be theoretically studied by first-principle (pQCD) calculations. The nature of such a medium can be deciphered only by confronting theoretical calculations with experimental data. 
Calculations beyond leading order (LO) in pQCD are of great importance to testing the reliability of theoretical studies and hence help improves our understanding of the properties of bulk matter. Significant progress has recently been made on this topic. And I shall only limit myself to a self-contained summary on the leading-logarithmic correction proportional to $\alpha_s \ln^2(L/l_0)$ obtained in a slightly generalized BDMPS-Z formalism\cite{Baier:1998,Zakharov:1996:1997}. Here, $L$ is the length of the QCD medium and $l_0$ is the size of its constituents. Such a correction is universally present in radiative $p_\perp$-broadening\cite{Wu:2011,Liou:Mueller:Wu:2013}, the transport coefficient $\hat{q}$\cite{Blaizot:2014,Iancu:2014} and radiative energy loss\cite{Blaizot:2014,Wu:2014,Arnold:2015}. Let us start with the discussion about the typical and averaged transverse momentum broadening at LO in \cite{Baier:1996,Arnold:2009}.

\section{Radiative $p_\perp$-broadening}

\subsection{Typical $p_\perp$-broadening at LO}\label{sec:typical}

$\hat{q}_R$ is defined as the typical transverse momentum squared per unit length transferred from the medium to a high energy parton in color representation $R$\cite{Baier:1996}. In terms of the gluon distribution $xG$ and the number density $\rho$ of the medium constituents, it can be written in the following form\cite{Baier:1996}
\begin{eqnarray}\label{eq:qhatDef}
\hat{q}_R\equiv \rho \int dq_\perp^2 q_\perp^2\frac{d\sigma_R}{dq_\perp^2}=\frac{4\pi^2\alpha_s C_R}{N_c^2-1} \rho xG(x,\hat{q}_R L)
\end{eqnarray}
with ${d\sigma_R}/{dq_\perp^2}$ the differential cross section for single scattering. 
%which holds in both hot and cold matter.
In a hot quark-gluon plasma (QGP) the gluon distribution in the wave function of each plasma particle is given by $xG_i(x,\hat{q}_R L)= \frac{\alpha_s C_i}{\pi} \ln\left(\frac{\hat{q}_R L}{m_D^2}\right)$ with $i=F$ for (anti-)quarks and $i=A$ for gluons and the (total) number density of each species is $\rho_i = 2 d_i n_i$ 
 with $d_i$ the dimension of color representation $i$. Here, $m_D$ ($\ll 1/L$) is the Debye mass and, in terms of the phase-space distribution $f^i$,%\footnote{It is also common to take $\epsilon_i=0$ in the literature (see, e.g., \cite{Arnold:2009}).},
\begin{eqnarray}
 n_i\equiv\int\frac{d^3p}{(2\pi)^3} f^i (1+\epsilon_i f^i)~~\mbox{with $\epsilon_{A/F}=\pm 1$}.
\end{eqnarray}
The following discussions shall be applicable to both cold and hot matter in terms of the corresponding $xG$.
 
A high-energy parton traversing a medium of length $L$ picks up a typical $p_\perp^2$ equal to $\hat{q}_R L$, in which only the contribution from multiple soft scattering is included\cite{Baier:1996}. This can be easily understood in terms of the mean free path of the high-energy parton (with $q_\perp$ a typical momentum transfer to the medium)%(in terms of the gluon distribution of the medium $\rho xG$)
\begin{eqnarray}
\frac{1}{\lambda_R} \equiv \sigma_R \rho xG \sim \frac{\alpha_s C_R}{q_\perp^2} \rho x G\sim\frac{\hat{q}_R}{q_\perp^2}.
\end{eqnarray}
By comparing $\lambda_R $ with $L$, one can distinguish typical multiple soft scattering with $q_\perp^2\lesssim \hat{q}_R L$ in each individual scattering from rare single hard scattering with $q_\perp^2\gg \hat{q}_R L$\cite{Arnold:2009}. The distribution of transverse momenta of the parton %, subject to uncorrelated multiple scattering, generically 
takes the form
\begin{eqnarray}\label{eq:Sdef}
\frac{dN}{d^2 p_\perp}%=\int\frac{d^2x_\perp}{(2\pi)^2}S(x)
=\int\frac{d^2x_\perp}{(2\pi)^2}\underbrace{e^{\rho L\int d^2q_\perp\frac{d\sigma_R}{d^2q_\perp}\left(e^{i{\bf p}_\perp\cdot{\bf x}_\perp}-1\right)}}_{S(x)}.
\end{eqnarray}
Including only typical multiple scattering gives, in the logarithmic approximation,
\begin{eqnarray}\label{eq:dNdp}
\frac{dN}{d^2p_\perp}\approx\frac{1}{\pi \hat{q}_R L}e^{-\frac{p_\perp^2}{\hat{q}_R L}}.
\end{eqnarray} 
And it is modified by rare hard single scattering only at $p_\perp^2 \gg \hat{q}_R L$ in which case one has
\begin{eqnarray}\label{eq:dNdpss}
\frac{dN}{d^2p_\perp}\approx\rho L \frac{d\sigma_R}{d^2p_\perp}\propto p_\perp^{-4}.
\end{eqnarray}
Quantities such as the average $p_\perp^2$ do receive large corrections from such rare scattering at high energies. However, the distribution $dN/d^2p_\perp$ carries more information %, is obviously more useful for theoretical calculations and more relevant for experimental observations. 
and its shape is mainly characterized by the typical $p_\perp^2$. This motivates us to focus only on multiple soft scattering and typical $p_\perp$-broadening shall be simply denoted by $\bra p_\perp^2\ket$ in all the following discussions.

\subsection{Double logarithmic correction to $p_\perp$-broadening}

In QCD matter the high-energy parton undergoes multiple scattering and radiates gluons. It hence generates a recoiling transverse momentum. Such a radiative correction to $\bra p_\perp^2 \ket$ has been studied in \cite{Wu:2011,Liou:Mueller:Wu:2013} using a formalism as an extension of that by BDMPS-Z\cite{Baier:1998,Zakharov:1996:1997}. The complete analytic evaluation of the contribution from one-gluon emission is complicated by the presence of multiple scattering but double logarithmic terms, $\ln^2(L/l_0)$, and single logarithmic terms, $\ln(L/l_0)$, can be evaluated analytically.%with $l_0$ the size of constituents of the medium\cite{Liou:Mueller:Wu:2013}.

The radiated gluon and the high-energy parton are in a coherent state within the formation time $t\equiv\frac{2 \omega}{k_\perp^2}$ with $\omega$ the gluon's energy and $k_\perp$ its transverse momentum. The evolution of the coherent pair is governed by a Schr\"{o}dinger-type evolution equation with a potential\cite{Wu:2011}
\begin{eqnarray}
V\approx -\frac{iN_c\hat{q}_R B_\perp^2}{4 C_R}~~\mbox{for $B_\perp^2\simeq\frac{1}{k_\perp^2}\gtrsim \frac{1}{\hat{q}_R L}$}.
\end{eqnarray}
The inverse of $V$ is the typical time scale for the pair of partons to undergo one individual collision and hence referred to as the mean free path of the coherent state
\begin{eqnarray}
\lambda_c\equiv \frac{4 C_R k_\perp^2}{N_c\hat{q}_R}\Rightarrow
\left\{
\begin{array}{ll}
\lambda_c \gtrsim t: & \mbox{single scattering},\\
\lambda_c \lesssim t: & \mbox{multiple scattering}.
\end{array}
\right.
\end{eqnarray}
By taking $\hat{q}_R\simeq\frac{m_D^2}{\lambda_R}$ one can see that $\lambda_c$ becomes larger than that of a single parton $\lambda_R$ when $k_\perp^2\gtrsim N_c m_D^2/(4 C_R)$.

The double logarithmic contribution only comes from a single scattering. As discussed above, the single scattering phase space can be singled out by requiring the formation time of the gluon, radiated inside of the medium, is not larger than the mean free path of the coherent pair, that is, 
\begin{eqnarray}\label{eq:dbRegion}
L\gtrsim \lambda_c\simeq\frac{k_\perp^2}{\hat{q}}\gtrsim t~~\mbox{and}~~L\gtrsim t \gtrsim l_0.
\end{eqnarray}
Within this region, the double logarithmic correction can be easily reproduced from the differential cross section for the emission process in single scattering
\bea
\left.\f{ d\sigma_{R\to R g}}{d\omega dk^2_\perp d^2p_{\perp}} =\f{ \alpha_s N_c }{ \pi^2 }\f{ q_\perp^2}{\omega k_\perp^2 p_\perp^2} \f{d \sigma_R}{ d^2q_{\perp}} \right|_{\vec{q}_\perp = \vec{k}_\perp + \vec{p}_{R\perp}}\label{equ:dIsrgdp2}
\eea
with the limits of the integral given by (\ref{eq:dbRegion}). And one has\cite{Liou:Mueller:Wu:2013}
\begin{eqnarray}\label{eq:pT2rad}
&&\Delta p_\perp^2(L)\equiv\frac{\alpha_s N_c}{\pi}\hat{q}_R L \int_{l_0}^L\frac{dt}{t}\int_{\hat{q}_R t}^{\hat{q}_R L}\frac{dk_\perp^2}{k_\perp^2}\nonumber\\
&&=\hat{q}_R L\Delta(L)~~\mbox{with $\Delta(L)\equiv\frac{\alpha_s N_c}{2\pi}\ln^2\left(\frac{L}{l_0}\right)$}.
\end{eqnarray}
Subleading single logarithmic terms have also been evaluated by crossing different boundaries between different regions in the phase space of the radiated gluon.

%That is, the distribution in transverse coordinate conjugate to $p_\perp$ is given by
%\begin{eqnarray}
%S(x_\perp)\approx 1-\frac{1}{4}(Q_s^2+\Delta p_\perp^2) x_\perp^2+O(x_\perp^4).
%\end{eqnarray} 

\begin{figure}
\begin{center}
\includegraphics[width=0.3\textwidth]{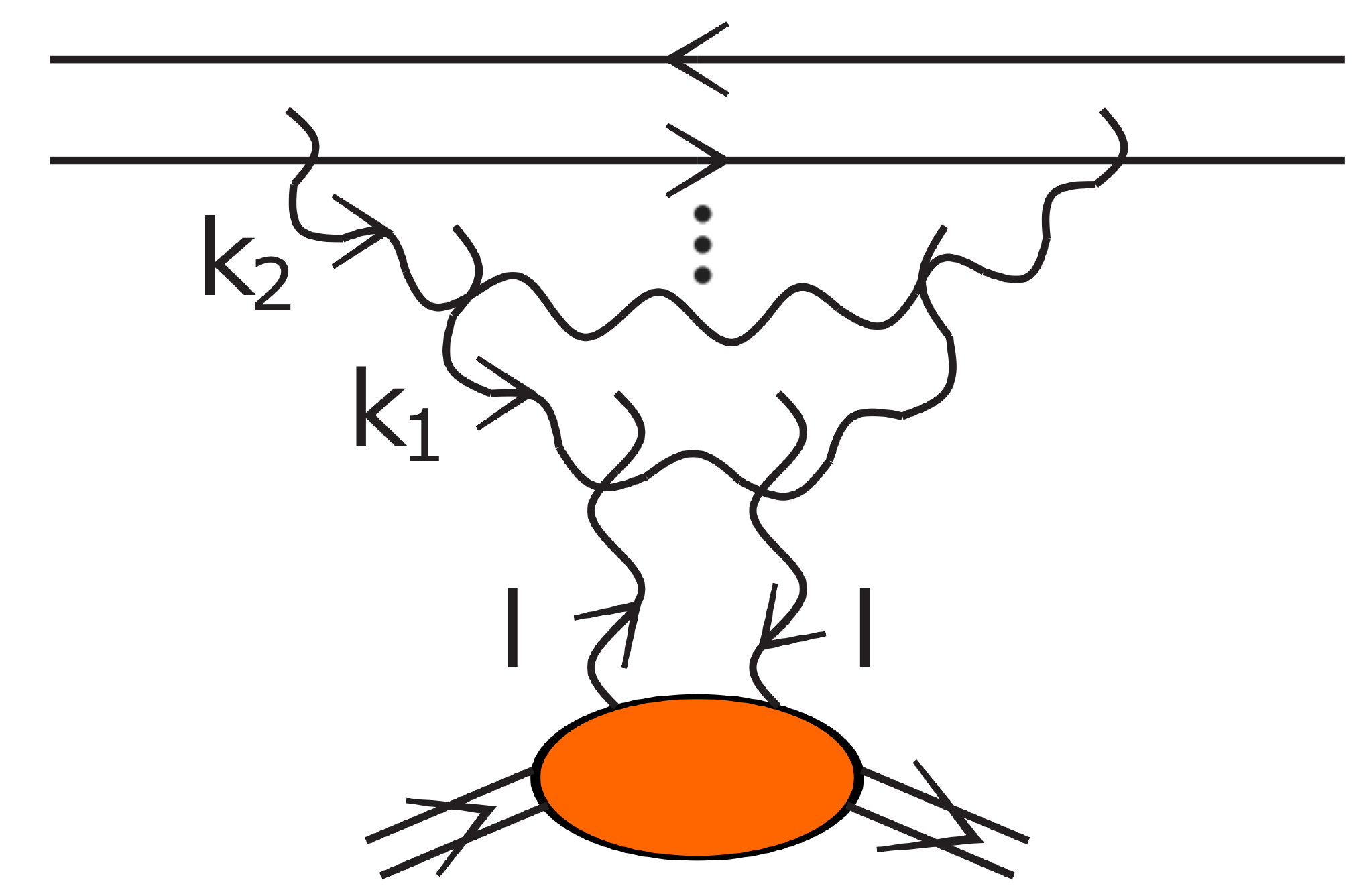}
\end{center}
\caption{Schematic illustration of diagrams for leading double logarithmic terms. The blob here denotes a constituent of the medium and $l$ denotes the momentum transfer between the constituent and the high-energy parton (dipole) with $n$ radiated gluons.% The double logarithmic region of integration over the gluon phase space corresponds to $t_1\lesssim t_2 \lesssim\cdots$ and $k_{1\perp}\lesssim k_{2\perp}\lesssim\cdots$.
}\label{fig:doubleb}
\end{figure}

A resummation of leading double logarithmic terms from radiating an arbitrary number of gluons induced by a single scattering can be easily carried out by repeating the calculation in (\ref{eq:pT2rad})\cite{Liou:Mueller:Wu:2013}. Let us start with two-gluon emission. As illustrated in Fig. \ref{fig:doubleb}, the term proportional to $\ln^4(L/l_0)$ comes only from the phase space in which the second gluon is radiated within the formation time of the first one and the transverse momentum of the second gluon is smaller than that of the first one. Accordingly, the next term in the leading double logarithmic series is given by
\begin{eqnarray}
&&\bra p_\perp^2\ket_2=\left(\frac{\alpha_s N_c}{\pi}\hat{q}_R L\right)^2\int_{l_0}^L\frac{dt_2}{t_2}\int_{\hat{q}_R t_2}^{\hat{q}_R L}\frac{dk_{2\perp}^2}{k_{2\perp}^2}\nonumber\\
&&\times\int_{l_0}^{t_1}\frac{dt_1}{t_1}\int_{\hat{q}_R t_1}^{k_{2\perp}^2}\frac{dk_{1\perp}^2}{k_{1\perp}^2}=\frac{(\hat{q}_RL)^2}{3!2!}[2\Delta(L)]^2.
\end{eqnarray}
By induction, one can show that the leading double logarithmic contribution from $n$-gluon emission is given by
\begin{eqnarray}\label{eq:pT2n}
\bra p_\perp^2\ket_n=\frac{(\hat{q}_RL)^n}{(n+1)!n!}[2\Delta(L)]^n
\end{eqnarray}
and hence obtain
\cite{Liou:Mueller:Wu:2013}.
\begin{eqnarray}\label{eq:pt2}
\bra p_\perp^2 \ket&=&\hat{q}_R L\left[2\Delta(L)\right]^{-\frac{1}{2}}I_1\left(\left[8\Delta(L)\right]^{\frac{1}{2}}\right).
\end{eqnarray}

The $p_\perp$-broadening in (\ref{eq:pT2rad}) is  the typical $p_\perp^2$ and, after being averaged over the path length, can be taken as an effective (renormalized) $\hat{q}_R$\cite{Blaizot:2014,Iancu:2014}
\begin{eqnarray}\label{eq:qhatren}
\hat{q}_R^{ren}(L)\approx\hat{q}_R\left[1+\Delta(L)\right].
\end{eqnarray}
Even though the correction includes contributions from radiating gluons with a life time $\sim L$, it can be still taken as a correction to the (local) transport coefficient as a consequence of the property of logarithmic integrations\cite{Blaizot:2014}. Let us only focus on leading double logarithmic terms from radiating $n$ gluons without overlapping formation times. They can be evaluated by repeating the calculation in (\ref{eq:pT2rad}) for each radiated gluon. Specifically, $S(x)$ receives a correction of the form 
\begin{eqnarray}
&&{\Delta S_n(x_\perp,L)}=-\frac{\alpha_s N_c}{4\pi n}{\hat{q}_R L  x_\perp^2}\int_{l_0}^Ld\ln t \ln\left(\frac{L}{t}\right)\nonumber\\
&&\times~\Delta S_{n-1}(x_\perp,L-t)
%\equiv \frac{\bar{\alpha}^n}{n!}\int_{l_0}^{L}d\ln t_1 \ln\left(\frac{L}{t_1}\right)\int_{l_0}^{L-t_1}d\ln t_2 \ln\left(\frac{L-t_1}{t_2}\right)\nn\\
%&&\times\int_{l_0}^{L-\sum\limits_{i=1}^{n-1}}d\ln t_n \ln\left(\frac{L-\sum\limits_{i=1}^{n-1}}{t_n}\right).
\end{eqnarray} 
with
\begin{eqnarray}
\Delta S_{1}(x_\perp,L)%&&=-{\bar\alpha}{4} \hat{q}_R L x_\perp^2\int_{l_0}^Ld\ln t \ln\left(\frac{L}{t}\right) e^{-\frac{1}{4}\hat{q}(L-t) x_\perp^2}\nn\\
\approx-\frac{1}{4} \hat{q}_R L\Delta(L) x_\perp^2 e^{-\frac{1}{4}\hat{q}_R L x_\perp^2}.
\end{eqnarray}
Each gluon has a formation time larger than $l_0$ and hence $\Delta S_{n}(x_\perp,L)$ vanishes if $L\leq n l_0$. This simply indicates that the double logarithmic approximation breaks down when $n\sim L/l_0$. Therefore, we have to restrict ourselves to the case $L\lesssim n l_0$. In this case leading logarithmic approximation applies and one has
\begin{eqnarray}
&&{\Delta S_n(x_\perp,L)}\approx\frac{e^{-\frac{1}{4}\hat{q}_R L x_\perp^2}}{n!}\left(-\frac{1}{4} \Delta p_\perp^2  x_\perp^2\right)^n.
%\equiv \frac{\bar{\alpha}^n}{n!}\int_{l_0}^{L}d\ln t_1 \ln\left(\frac{L}{t_1}\right)\int_{l_0}^{L-t_1}d\ln t_2 \ln\left(\frac{L-t_1}{t_2}\right)\nn\\
%&&\times\int_{l_0}^{L-\sum\limits_{i=1}^{n-1}}d\ln t_n \ln\left(\frac{L-\sum\limits_{i=1}^{n-1}}{t_n}\right).
\end{eqnarray}
From this, one can justify the exponentiation of the double logarithmic integral in (\ref{eq:pT2rad})
\begin{eqnarray}\label{eq:S}
S(x_\perp)=e^{-\frac{1}{4}\hat{q}_R L x_\perp^2}\sum\limits_{n\lesssim L/l_0}\Delta S_n\approx e^{-\frac{1}{4}(\hat{q}_R L+ \Delta p_\perp^2)x_\perp^2},
\end{eqnarray}
if $x_\perp$ satisfies 
\begin{eqnarray}\label{eq:x2Cond}
\frac{(\frac{1}{4} \Delta p_\perp^2x_\perp^2)^{L/l_0}}{(L/l_0)!}< e^{-\frac{1}{4} \Delta p_\perp^2 x_\perp^2}.
\end{eqnarray}
At $L\gg l_0$ it asymptotically gives $x_\perp^2\lesssim \frac{4L}{l_0 \Delta p_\perp^2}\gg\frac{4}{\Delta p_\perp^2}$ and, in this case, plugging (\ref{eq:S}) into (\ref{eq:Sdef}) gives
\begin{eqnarray}
\frac{dN}{dp_\perp^2}\approx \frac{e^{-\frac{p_\perp^2}{\hat{q}_R L (1+\Delta(L))}}}{\hat{q}_R L (1+\Delta(L))}.
\end{eqnarray}
This justifies that the leading logarithmic result in (\ref{eq:pT2rad}) is the typical value of $p_\perp^2$.

\section{Radiative parton energy loss}

At high energies, the parton loses its energy mainly due to medium-induced gluon radiation (the LPM effect). Since the formation time of a radiated gluon grows with its energy, the average energy loss within any period of time $t\gtrsim \lambda_R$ is dominated by radiating one gluon with the formation time comparable with $t$. Within $t\simeq {2\omega}/{k_\perp^2}$, the gluon picks up a transverse momentum broadening $k_\perp^2 = \hat{q} t$. Accordingly, one has $t\simeq \sqrt{{2\omega}/{\hat{q}}}$ and the maximum energy that the gluon may carry away is given by $\omega_c(t)=\hat{q} t^2$. From this, one can get parametrically two well-known results: the LPM spectrum\cite{Baier:1998,Zakharov:1996:1997}
\begin{eqnarray}\label{eq:dIdo}
\omega\frac{dI}{d\omega}\sim \alpha_s N_c \frac{L}{t(\omega)}\sim\alpha_s N_c \sqrt{\frac{\hat{q}L}{\omega}}
\end{eqnarray}
and the average energy loss per unit length
\cite{Baier:1996}
\be\label{eq:dept2}
-\f{dE}{dz} \sim \alpha_s N_c \f{\omega_c(L)}{L}=\alpha_s N_c \hat{q}L.
\ee
Will these LO results be modified by the radiative correction? The answer is: in double logarithmic accuracy the only thing one needs to do is replace $\hat{q}$ by $\hat{q}^{ren}$ in (\ref{eq:qhatren}), as justified in \cite{Blaizot:2014,Iancu:2014}. This has been further confirmed by a detailed calculation of average energy loss from radiating two and more gluons\cite{Wu:2014} and a more elaborated and careful analysis of two-gluon emission beyond the double logarithmic approximation\cite{Arnold:2015}. In the following I shall only take average energy loss for example to illustrate the underlying physical picture.

The energy loss due to two-gluon emission $\Delta E_2$ is greatly simplified in the double logarithmic region discussed in the previous section. The first gluon with energy $\omega_1$, similar to that in the case with one-gluon emission, typically has a formation time $t_1 \simeq L$, $\omega_1\simeq \hat{q} L^2$ and $k_{1\perp}^2 \simeq {\hat{q} L}$. In the double logarithmic region, the second gluon with energy $\omega_2$ typically has  
\be
k_{2\perp}^2 \lesssim k_{1\perp}^2,~~\omega_2 \lesssim \omega_1,~~t_2\lesssim \sqrt{\f{\omega_2}{\hat{q}_A}}\lesssim\sqrt{\f{\omega_1}{\hat{q}_A}}\simeq t_1.\ee 
Therefore, the second gluon plays the same role as that in the calculation of the radiative $p_\perp$-broadening in (\ref{eq:pT2rad}). After integrating out the contribution from the second gluon, we have\cite{Wu:2014}
\bea\label{eq:DE2log}
\Delta E_2%\simeq \f{\alpha_s N_c}{4 \pi}\hat{q}\text{Re}~i \int \f{d\omega_1}{\omega_1}\int_0^L dt_1 (L-t_1) \Delta(t_1)\nonumber\\
%\times \f{\Omega_1 \sin(\Omega_1 t_1) - \Omega_1 t_1 \cos(\Omega_1 t_1)}{\sin^3(\Omega_1 t_1)}
\approx\f{\alpha_s N_c}{12}\hat{q}L^2 \Delta(L),
\eea
which has exactly the same prefactor as the LO result (for the parton approaching the medium from outside) in\cite{Baier:1998}, i.e.,
$\Delta E_1 = \f{\alpha_s N_c}{12} \hat{q} L^2$.

The resummation of the double logarithmic correction in eq. (\ref{eq:DE2log}) can be carried out in exactly the same way as that in the calculation of $\bra p_\perp^2\ket$. For $(n+1)$-gluon emission, besides the most long-lived gluon, the other $n$ gluons undergoes one single scattering. Using the same orderings in the formation time and transverse momenta of these $n$ gluons that lead to (\ref{eq:pT2n}) gives a contribution of the form
\bea\label{eq:DE2logResum}
\Delta E_{n+1}=\f{\alpha_s N_c}{12}L\f{\hat{q} L}{n!(n+1)!}\left[\f{\alpha_s N_c}{\pi}\ln^{2}\f{L}{l_0}\right]^n.\eea
Therefore, the total energy loss is given by\cite{Wu:2014}
\bea
\Delta E = \sum\limits_{n=0}^\infty\Delta E_{n+1} = \f{\alpha_s N_c}{12}L\bra p_\perp^2\ket
\eea
with $\bra p_\perp^2\ket$ given by (\ref{eq:pt2}), including all the leading double logarithmic contributions from radiating arbitrary number of gluons.

\section{Discussions}

Our discussions so far have been focused on the medium-induced radiative corrections. Rather, in the above results the limit $\hat{q}\to 0$ has been subtracted out\cite{Baier:1998,Zakharov:1996:1997}. One effect neglected from such a subtraction is the running of the QCD coupling constant. We have also discussed this effect in the radiative correction to $p_\perp^2$ from one-gluon emission and find a factor 2 larger than in the fixed coupling case\cite{Liou:Mueller:Wu:2013}. How the inclusion of the running coupling modifies the evolution of $\hat{q}^{ren}$ has also been discussed in more detail in \cite{Iancu:Trian:2014}.

At the end, I briefly comment on the differences between our double logarithmic radiative correction to $p_\perp^2$ and some previous results beyond LO in \cite{CaronHuot:2008, CasalderreySolana:Wang:2007}. In \cite{CaronHuot:2008}, the NLO ($O(g)$) correction to transverse momentum broadening has been obtained analytically in a hot ($g\ll 1$) quark-gluon plasma using a perturbative-kinetic approach. This result was calculated from the soft contribution to the two-body collisional kernel from soft collisions with $q_\perp\sim gT$. In contrast our result comes from radiation, which is of $O(g^2)$ and would be present at next-next-to-leading order in that approach\cite{Ghiglieri:2015}. 

The authors of Ref. \cite{CasalderreySolana:Wang:2007} (see \cite{Kang:2015} for a recent progress) have also calculated the radiative correction to $p_\perp$-broadening in a QGP. The effect of multiple (soft) scattering was not considered in this formalism. As a result, the regime of double logarithmic approximation (Sec.~V) has been extended into $Q^2>k_\perp^2> \mu^2$ and $E/(\hat{q}L)>t> l_0$. This is different from ours in (\ref{eq:dbRegion}) in which the phase space includes only single scattering inside the medium (multiple scattering sets the boundary at $k_\perp^2\simeq\hat{q}t$). %, which has not been considered in our calculation since we would like to emphasize the equivalence between $\hat{q}$ and typical $p_\perp^2$. 
It will be intriguing to see how to formulate multiple scattering in this formalism.


%% References

\nocite{*}
\bibliographystyle{elsarticle-num}
\bibliography{martin}


\end{document}